\documentclass[twoside,english]{elsarticle}
\usepackage[T1]{fontenc}
\usepackage[latin9]{inputenc}
\usepackage{amsmath}
\usepackage{graphicx}

\makeatletter

\providecommand{\tabularnewline}{\\}

\journal{ArXiv}

\makeatother

\usepackage{babel}
\begin{document}

\begin{frontmatter}{}

\title{Causal unitary qubit model of black hole evaporation}

\author{Bogus\l aw Broda}

\ead{boguslaw.broda@uni.lodz.pl}

\ead[url]{http://merlin.phys.uni.lodz.pl/BBroda}

\address{Department of Theoretical Physics, Faculty of Physics and Applied
Informatics, University of \L ód\'{z}, 90-236 \L ód\'{z}, Pomorska
149/153, Poland}
\begin{abstract}
A new simple qubit model of black hole evaporation is proposed. The
model operates on four qubits and is defined in terms of quantum gates
and a quantum circuit. The chief features of the model include explicit
unitarity and (most notably) causality which is understood as the
impossibility of the transfer of information from the interior of
the black hole through its horizon. The corresponding von Neumann
entanglement entropy yields a crude (four-qubit) approximation of
the Page curve.
\end{abstract}
\begin{keyword}
black hole information loss paradox \sep qubit model of black hole
evaporation \sep causal and unitary black hole evaporation \sep
von Neumann entanglement entropy \sep the Page curve

\PACS 04.70.Dy \sep 03.67.Ac \sep 04.60.-m

\MSC[2020]83C57 \sep 81P65

\end{keyword}

\end{frontmatter}{}

\section{Introduction}

\begin{figure}[h]
\begin{centering}
\begin{tabular}{cc}
\includegraphics[scale=0.45]{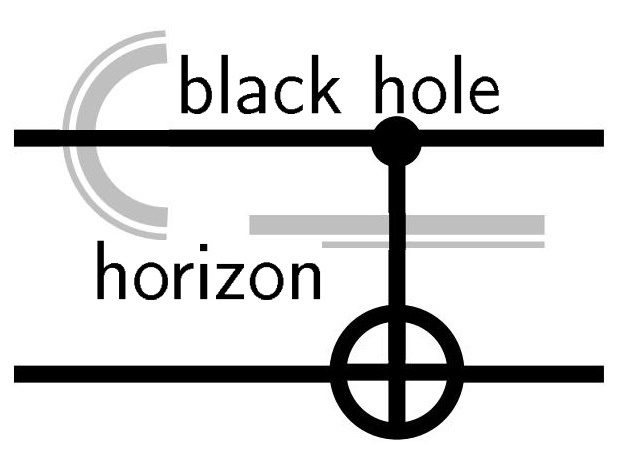} & \includegraphics[scale=0.45]{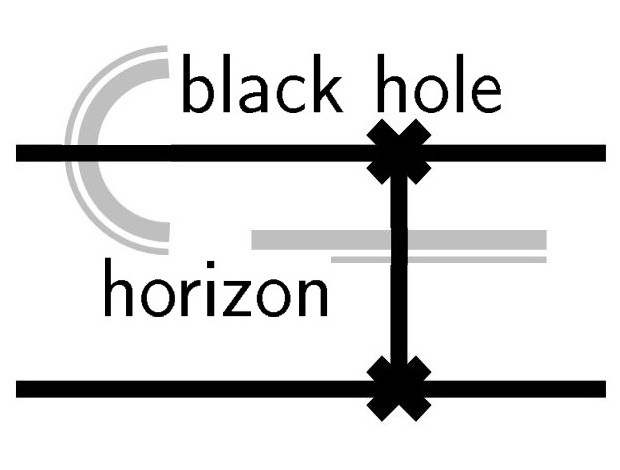}\tabularnewline
\includegraphics[scale=0.45]{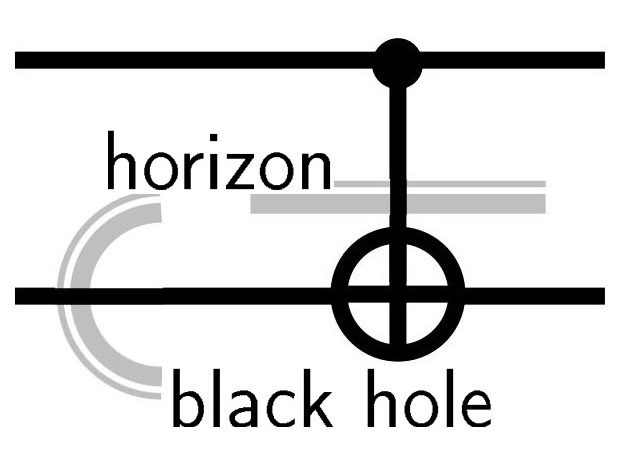} & \includegraphics[scale=0.45]{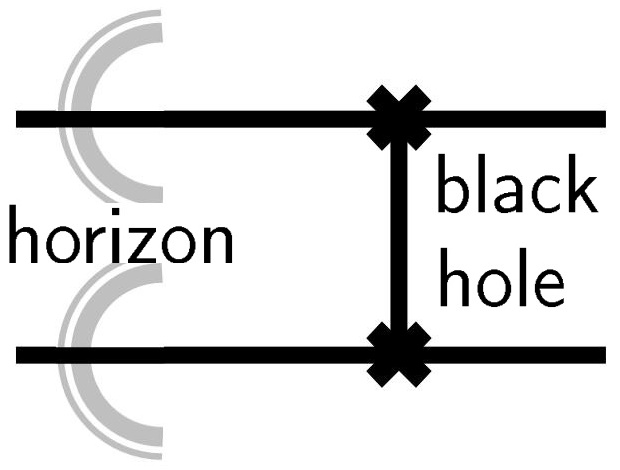}\tabularnewline
\end{tabular}
\par\end{centering}
\caption{Four examples of quantum gates and their specific locations in the
context of the presence of the BH horizon. The two left panels present
the CNOT gates, whereas the two right ones, the SWAP gates. In our
approach, the two upper gates are causally forbidden in their proposed
locations, whereas the two lower gates are causally allowed. The double
gray line denotes the BH horizon, with its thicker component corresponding
to the inner part of the BH.}

\centering{}\label{Fig:1}
\end{figure}

The black hole (BH) information (loss) paradox is a problem concerning
difficulties with the unitarity of the process of BH evaporation and
evolution (see e.g.\ \citet{Hawking1976}, or reviews \citet{Chakraborty2017,Harlow2016,Polchinski2017,Marolf2017}).
The assumption (ours, in particular) that the unitarity is conserved
implies several general scenarios. For example, one can adopt the
scenario (as also we do) that information is being (somehow) gradually
released during BH evaporation. However this point of view (as obviously
any other) requires an indication of some convincing physical mechanism,
or (in the case of the lack of thereof) at least some working abstract
algorithm for the transfer of information. One of the obvious approaches
to study the paradox, abstracting from a particular physical mechanism,
consists in an analysis of the problem in terms of qubits. In literature,
we can find numerous qubit models more or less successfully reproducing
various steps involved in BH evolution (see e.g.\ \citet{Broda2020,Broda2021,Giddings2012,Giddings2012x,Giddings2013,Mathur2009,Mathur2009x,Mathur2011,Osuga2018},
or review \citet{Avery2013}). Unfortunately, it seems that in all
these models the important issue of causality has not attracted due
attention, and consequently the possibility of faster-than-light communication
has not been explicitly excluded. In contrast to this is our present
treatment, where causality is a priority. More precisely, in our approach,
we impose strict control on the direction of information transfer
through the BH horizon. As a (sufficient) minimum, we assume that
any transfer of information from a BH through its horizon is forbidden.
On the other hand, any transfer of information through the horizon
in the opposite direction (from outside to inside), as well as between
particles (qubits) residing exclusively outside or inside the BH is
not restricted. Fortunately, working in terms of qubits (and consequently,
in terms of quantum gates and a quantum circuit) gives us an excellent
possibility to control the direction of the transfer of information.
In particular, the CNOT (controlled NOT) gate acting between qubits
on both sides of the BH horizon is only allowed provided the control
qubit is exactly the outer qubit, and the target qubit is the inner
one. Clearly, the SWAP gate in this position (acting across the horizon)
is forbidden. However, all kinds of gates acting exclusively outside
or inside the BH are allowed (see Fig.~\ref{Fig:1}).

The scenario with the gradual release of information during BH evaporation
suggests the following subscenarios:

\emph{Information passes through the BH horizon.} In principle, this
is the most obvious possibility, but in consequence information seems
to be permanently locked inside the BH (and possibly also lost), or
faster-than-light communication should be invoked for its release
(information cannot escape the BH horizon without traveling faster
than light). So, we could ask after \citet{Polchinski2017}: ``How
does information travel from inside the black hole to the outside?''

\emph{Information does not pass through the BH horizon.} This subscenario
evidently solves the problem with information retrieval, but it creates
problems with the BH itself. Apparently, the BH does not appear in
this subscenario, or it is not being populated by matter.

\emph{Information is both reflected at the BH horizon and passes through
the horizon.} On first sight, the above difficulties with information
retrieval and the BH itself seem to be solved. Unfortunately, the
existence of two copies of information (the first copy inside the
BH and the second one outside the BH) requires convincing arguments
that we are not conflicted with the no-cloning theorem (and at later
stages, possibly also, with the no-deleting theorem). For the corresponding
discussion in the framework of the, so-called, BH complementarity
see e.g. \citet{Susskind2004}.

\emph{Information resides in the entangled states of (pairs of) qubits
on both sides of the BH horizon}. This is the point of view assumed
in this paper.

\section{The four-qubit model}

The model is defined in a finite dimensional Hilbert space $\mathcal{H}$,
in the language of qubits. To automatically ensure unitarity of the
model, the quantum evolution is described in terms of quantum gates
operating in the framework of a quantum circuit. We have fixed four
qubits as a minimal number of qubits necessary to demonstrate the
proposed mechanism (a huge tensor power of the four-qubit block could
be considered as an approximation to a possible fuller model). Each
of the four qubits has assigned a specific task:

$1^{\textrm{st}}$ qubit, denoted by $m$ ($m$ like ``matter''),
is the ``original'' information carrying qubit (one of those) initially
forming the BH or passing through the BH horizon of the already formed
BH. 

$2^{\textrm{nd}}$ qubit, denoted by $g$ ($g$ like ``graviton''),
is an auxiliary qubit for some time being entangled with the qubit
$m$. Its principal task consists in collective transport (together
with the qubit $m$) of the information initially contained in $m$
without violating the no-cloning theorem (and at later stages, possibly
also, the no-deleting theorem).

$3^{\textrm{rd}}$ qubit, denoted by ``$-$'' (the minus sign),
corresponds to a Hawking particle with the negative energy $-\omega$,
which contributes to the negative energy flux required for BH evaporation.

$4^{\textrm{th}}$ qubit, denoted by ``$+$'' (the plus sign), corresponds
to a Hawking particle with the positive energy $\omega$, which compensates
(according to energy conservation) negative energy of the ``$-$''
particle.

\begin{figure}[h]
\begin{centering}
\includegraphics[scale=0.45]{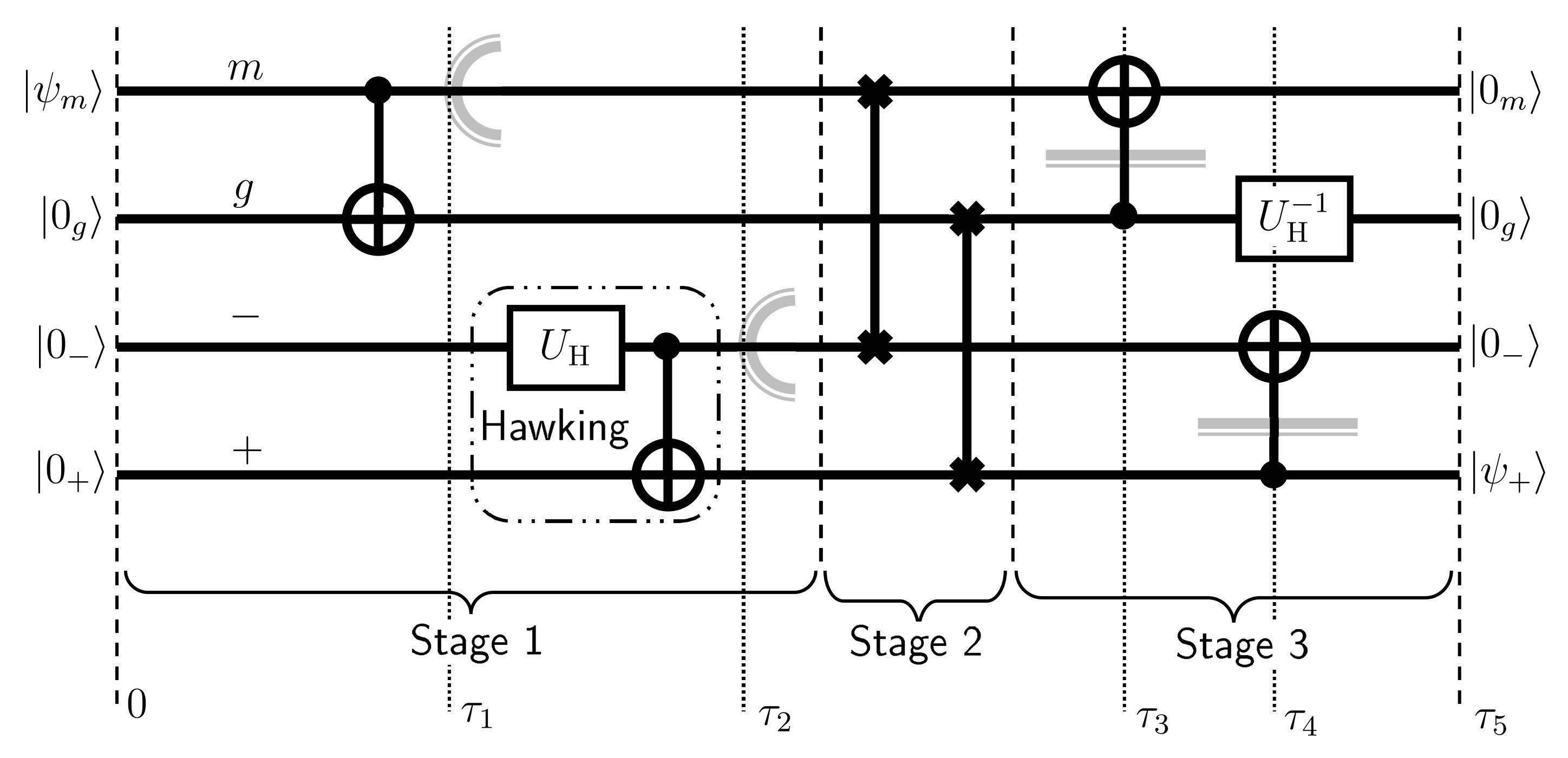}
\par\end{centering}
\centering{}\caption{The quantum circuit performing the algorithm proposed in the paper.
The algorithm acts on four qubits with precisely assigned tasks (described
in the text). The (four) vertical dashed lines divide the algorithm
onto three Stages. The (four) vertical dotted lines correspond to
abrupt changes of the von Neumann entanglement entropy at the instants
$\tau_{1}$, $\tau_{2}$, $\tau_{3}$ and $\tau_{4}$. The box labeled
``Hawking'' contains the gates responsible for the production of
Hawking pairs. The double gray lines are described in Fig.~\ref{Fig:1}}
\label{Fig:2}
\end{figure}

The quantum circuit in Fig.~\ref{Fig:2} performs the algorithm proposed
in the paper, and it works as follows. Initially, the qubit $m$ (``matter'')
is in the arbitrary but fixed state

\begin{equation}
\left|\psi_{m}\right\rangle =\lambda\left|0_{m}\right\rangle +\mu\left|1_{m}\right\rangle ,\label{eq:psi_m}
\end{equation}
whereas the rest of the qubits, $g$, ``$-$'', ``$+$'', is in
the ``vacuum'' state
\begin{equation}
\left|0_{g}\right\rangle ,\quad\left|0_{-}\right\rangle ,\quad\left|0_{+}\right\rangle ,\label{eq:vacuum_g_minus_plus}
\end{equation}
respectively. Nowhere do we refer to the formalism of the (antisymmetric)
Fock space, but our physics picture of qubits is close to a fermionic
description of matter. Hence, in our terminology the state $\left|0\right\rangle $
is called the vacuum state, and so interpreted. The total initial
state is then
\begin{equation}
\left|\varPsi_{mg-+}^{0}\right\rangle =\left|\psi_{m}\right\rangle \left|0_{g}0_{-}0_{+}\right\rangle \equiv\left(\lambda\left|0_{m}\right\rangle +\mu\left|1_{m}\right\rangle \right)\left|0_{g}0_{-}0_{+}\right\rangle .\label{eq:psi0_m_g_minus_plus}
\end{equation}

Next a short series of quantum operations (gates) acts independently
in the first pair of the qubits ($m$ and $g$), and in the second
one (``$-$'' and ``$+$''), respectively, as described in the
framework of Stage~1 in Fig.~\ref{Fig:2}. As a result, the quantum
information initially contained only in the qubit $m$ is now contained
in the following state of the first pair of the entangled qubits ($m$
and $g$),
\begin{equation}
\left|\phi_{mg}\right\rangle =\lambda\left|0_{m}0_{g}\right\rangle +\mu\left|1_{m}1_{g}\right\rangle .\label{eq:phi_m_g}
\end{equation}
In turn, the evolution of the second pair of the qubits (``$-$''
and ``$+$'') yields a Hawking pair given by the state
\begin{equation}
\left|\chi_{-+}\right\rangle =\alpha\left|0_{-}0_{+}\right\rangle +\beta\left|1_{-}1_{+}\right\rangle .\label{eq:chi_minus_plus}
\end{equation}
The symbol ``$U_{\textrm{H}}$'' in Fig.~\ref{Fig:2} denotes a
one-qubit unitary transformation ($U_{\textrm{H}}^{-1}$ is its inverse)
defined by (or defining) the Hawking process specified by the complex
coefficients $\alpha$ and $\beta$. Finally, the total state of the
system produced in the framework of Stage 1 is 
\begin{equation}
\left|\varPsi_{mg-+}^{1}\right\rangle =\left|\phi_{mg}\right\rangle \left|\chi_{-+}\right\rangle \equiv\left(\lambda\left|0_{m}0_{g}\right\rangle +\mu\left|1_{m}1_{g}\right\rangle \right)\left(\alpha\left|0_{-}0_{+}\right\rangle +\beta\left|1_{-}1_{+}\right\rangle \right).\label{eq:psi1_m_g_minus_plus}
\end{equation}

Once the qubit $m$, and (independently) ``$-$'', passes through
the BH horizon, a corresponding contribution to the von Neumann entanglement
entropy increases by
\begin{equation}
S'=-\left|\lambda\right|^{2}\ln\left|\lambda\right|^{2}-\left|\mu\right|^{2}\ln\left|\mu\right|^{2},\label{eq:s_prime}
\end{equation}
and

\begin{equation}
S''=-\left|\alpha\right|^{2}\ln\left|\alpha\right|^{2}-\left|\beta\right|^{2}\ln\left|\beta\right|^{2},\label{eq:s_bis}
\end{equation}
respectively. We could interpret $S'$, coming from the first pair
of qubits ($m$ and $g$), as the Bekenstein\textendash Hawking entropy
of the BH due to its matter content, and $S''$, coming from the second
pair (``$-$'' and ``$+$''), as entanglement entropy due to Hawking
radiation, respectively. The total four-qubit contribution to entropy
is then $S_{1}=S'+S''$.

\begin{figure}[h]
\begin{centering}
\begin{tabular}{c}
(a) \includegraphics[viewport=0bp 0bp 337bp 140bp,clip,scale=0.45]{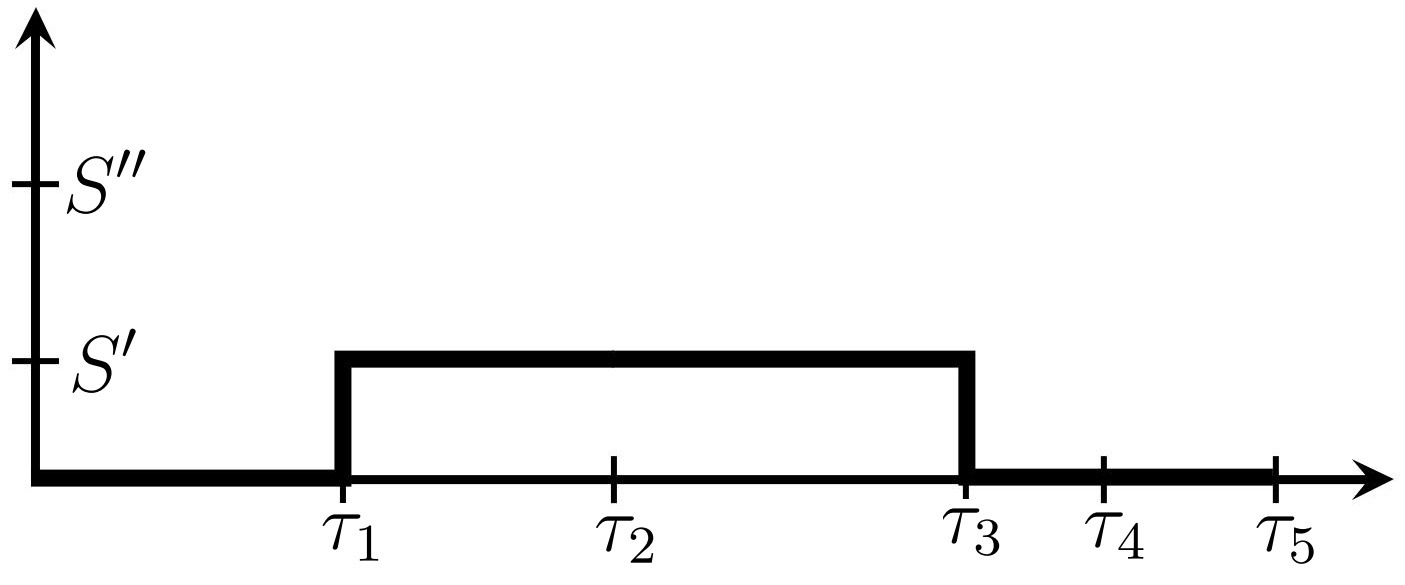}\tabularnewline
(b) \includegraphics[scale=0.45]{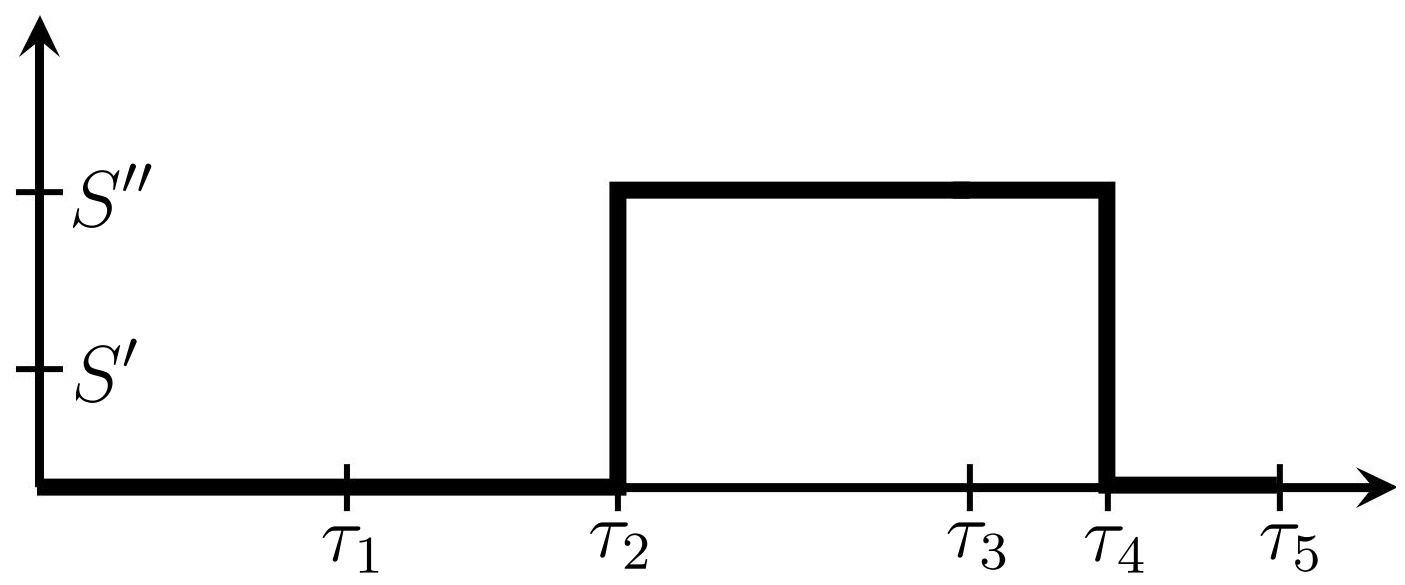}\tabularnewline
(c) \includegraphics[scale=0.45]{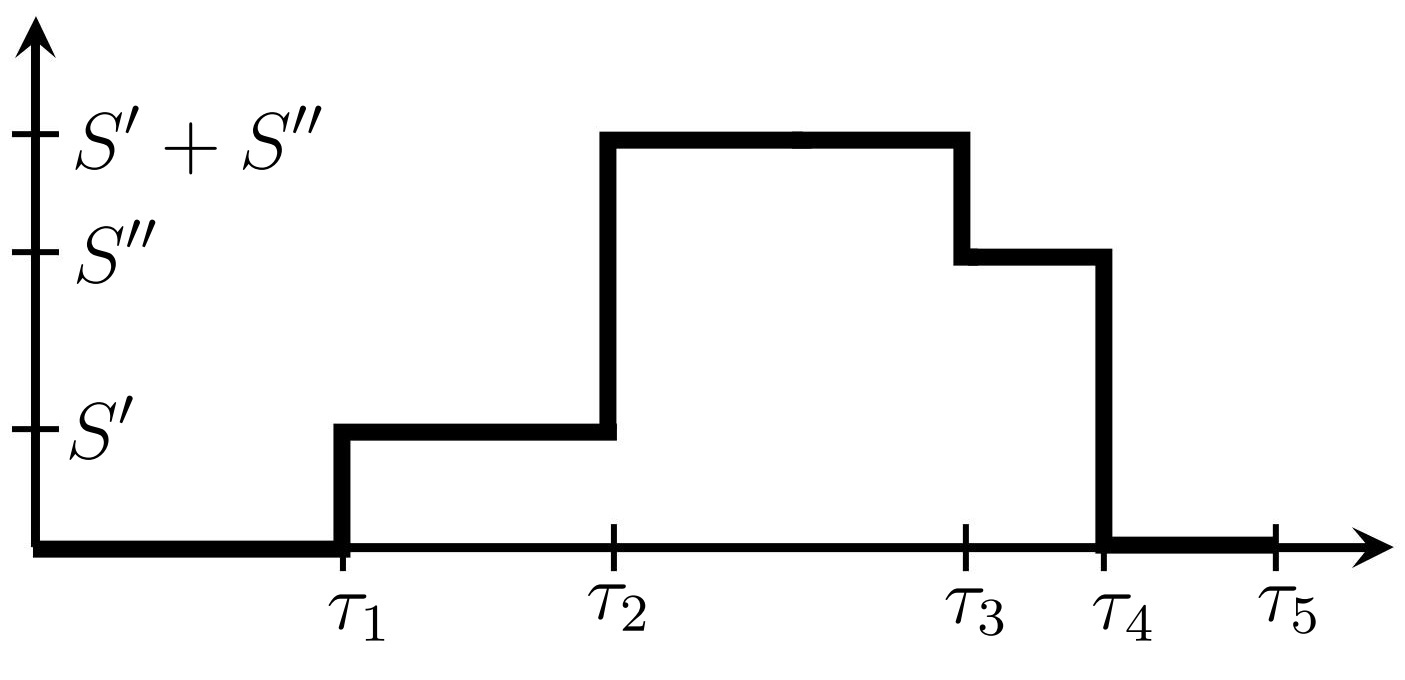}\tabularnewline
\end{tabular}
\par\end{centering}
\caption{A sample four-qubit ``Page curve''. The upper (a) and the middle
(b) panel presents a sample contribution to the von Neumann entanglement
entropy from the first ($m$ and $g$) and the second (``$-$''
and ``$+$'') qubit pair, respectively. The lower panel (c) presents
their sum, i.e.\ the total value of the von Neumann entanglement
entropy. The instants $\tau_{1}$, $\tau_{2}$, $\tau_{3}$ and $\tau_{4}$
are defined in Fig.~\ref{Fig:2}.}

\centering{}\label{Fig:3}
\end{figure}

For graphical definiteness, we have assumed a particular order of
elements in Fig.~\ref{Fig:2}. However, some flexibility in their
order is allowed. Actually, positions of the elements can be (continuously)
deformed, at least within each of the three Stages (i.e.\ between
vertical dashed lines), without any consequences for our discussion.

Stage~2 in Fig.~\ref{Fig:2}, containing two SWAP operations, one
entirely inside the BH and one entirely outside the BH, yields the
total state
\begin{equation}
\left|\varPsi_{mg-+}^{2}\right\rangle =\left(\alpha\left|0_{m}0_{g}\right\rangle +\beta\left|1_{m}1_{g}\right\rangle \right)\left(\lambda\left|0_{-}0_{+}\right\rangle +\mu\left|1_{-}1_{+}\right\rangle \right).\label{eq:psi2_m_g_minus_plus}
\end{equation}
Obviously the entropy is left unchanged, i.e.\ $S_{2}=S_{1}$. The
(final) Stage~3 completes the information transfer, and the total
state assumes the form
\begin{equation}
\left|\varPsi_{mg-+}^{3}\right\rangle =\left|0_{m}0_{g}0_{-}\right\rangle \left(\lambda\left|0_{+}\right\rangle +\mu\left|1_{+}\right\rangle \right)\equiv\left|0_{m}0_{g}0_{-}\right\rangle \left|\psi_{+}\right\rangle .\label{eq:psi3_m_g_minus_plus}
\end{equation}
The total entropy is now $S_{3}=0$.

It is interesting to analyze our model from the perspective of the
Page curve (see \citet{Page1993}). Since the model consists of only
four qubits we should not expect to much and require the emergence
of a genuine Page curve. For our four-qubit model a sample ``Page
curve'' is presented in Fig.~\ref{Fig:3}c. The entropy changes
(abruptly) only at the instants $\tau_{1}$, $\tau_{2}$, $\tau_{3}$
and $\tau_{4}$ corresponding to crossings of the BH horizon (see
Fig.~\ref{Fig:2}). The staircase-shape curve in Fig.~\ref{Fig:3}c
can be interpreted as a crude approximation of the genuine smooth
tent-shape Page curve. 

\section{Summary}

We have proposed a new simple four-qubit model of BH evaporation,
formulated in terms of quantum gates operating in the framework of
a quantum circuit. The unitarity of the model is automatically satisfied
by virtue of the gate construction. Moreover, most notably, the model
is causal, which is understood as the impossibility of the transfer
of information from the interior of the BH through its horizon. The
causality is built in the model by appropriate selection of gates
and their positions with respect to the BH horizon (see Fig.~\ref{Fig:1}
and the accompanying discussion). Thus, the model unitarily and causally
transfers quantum information from the ``original'' ingoing ``matter''
qubit $m$ to the outgoing Hawking qubit ``$+$''.

The corresponding von Neumann entanglement entropy yields a crude
(four-qubit) staircase-shape approximation (see Fig.~\ref{Fig:3}c)
of the Page curve. Alternatively, one can consider only the part $S''$
of entropy (coming from the Hawking radiation) as a contribution to
the Page curve. This even seems to be more in the original spirit
of Page, and moreover qualitatively changes nothing in our reasoning.
The step-shape curve in Fig.~\ref{Fig:3}b is a crude approximation
of the Page curve as well. One could further argue that a possible
huge tensor power of the four-qubit block (interpreted as an approximation
to a hypothetical fuller model with a huge number of qubits) could
yield a smooth curve as a result of the summation of the huge number
of the four-qubit staircase-shape curves, or step-shape ones, corresponding
to Fig.~\ref{Fig:3}c, or Fig.~\ref{Fig:3}b, respectively, with
the sets of instants $\tau_{1},$ $\tau_{2},$ $\tau_{3},$ $\tau_{4}$
a bit different for each independent four-qubit block.

\bibliographystyle{elsarticle-harv}
\bibliography{bhinfopar}

\end{document}